\documentclass[aps,prl,twocolumn, superscriptaddress,nofootinbib,groupedaddress]{revtex4}  % for review and submission

\usepackage{graphicx}
\usepackage{epsfig}
\usepackage{bm}
\usepackage{amsmath}
\usepackage{amssymb}
\usepackage{wasysym}
\usepackage{epstopdf}
\usepackage{color}
\usepackage{xcolor}
\usepackage{epstopdf}
\usepackage{upgreek}
\usepackage{natbib}
\usepackage{arydshln}
\usepackage[T1]{fontenc}

\usepackage{natbib}

\usepackage{psfrag}
\usepackage{color}
\usepackage{graphicx,float,amssymb}
\usepackage{color}
\usepackage{graphicx}
\usepackage{natbib}
\usepackage{bm}
\usepackage{amsmath}

\newcommand{\bfx}{{\bf x}}

\newcommand{\bfu}{{\bf u}}

\newcommand{\bfX}{{\bf X}}
\newcommand{\bfU}{{\bf U}}
\newcommand{\bfe}{{\bf e}}

\usepackage[colorlinks=false, bookmarks=true]{hyperref}

\usepackage{graphicx}
\usepackage{epsfig}
\usepackage{bm}
\usepackage{amsmath}
\usepackage{amssymb}
\usepackage{wasysym}
\usepackage{epstopdf}
\usepackage{color}
\usepackage{xcolor}
\usepackage{epstopdf}
\usepackage{upgreek}
\usepackage{natbib}
\usepackage{arydshln}
\usepackage[T1]{fontenc}

\begin{document}

%\preprint{AIP/123-QED}
%\title{Gravity-Driven Accelerations of Small Buoyant Particles in Turbulence}
%\title{Gravity-Driven Accelerations of Small Particles in Turbulence}
%\title{Tiny Particles do not Always Trace the Turbulent Flow}
%\title{Why Buoyancy is Prohibitive when Using Microbubbles and Microparticles \\ as Tracers in Turbulence}
\title{Micro-bubbles and Micro-particles are Not Faithful Tracers of Turbulent Acceleration}

\author{Varghese Mathai}
\affiliation{Physics of Fluids Group, Faculty of Science and Technology,
Mesa+ Institute, University of Twente, 7500 AE Enschede, The Netherlands.}

\author{Enrico Calzavarini}
\thanks{enrico.calzavarini@polytech-lille.fr}
\affiliation{ 
Univ. Lille, CNRS, FRE 3723, LML, Laboratoire de Mecanique de Lille, F 59000 Lille, France}
%Laboratoire de M\'ecanique de Lille CNRS/UMR 8107, Universit\'e Lille 1 et \'Ecole Polytechnique Universitaire de Lille, Cit\'e Scientifique Avenue Paul Langevin, 59650 Villeneuve d'Ascq, France.}

\author{Jon Brons}
\affiliation{Physics of Fluids Group, Faculty of Science and Technology,
Mesa+ Institute, University of Twente, 7500 AE Enschede, The Netherlands.}
\affiliation{ Department of Mathematics and Physics, Faculty of Engineering and Computing, Coventry University, United Kingdom}

\author{Chao Sun}
\thanks{chaosun@tsinghua.edu.cn}
\affiliation{Center for Combustion Energy and Department of Thermal Engineering, 
Tsinghua University, 100084 Beijing, China.}
\affiliation{Physics of Fluids Group, Faculty of Science and Technology,
Mesa+ Institute, University of Twente, 7500 AE Enschede, The Netherlands.}

\author{Detlef Lohse}
\affiliation{Physics of Fluids Group, Faculty of Science and Technology,
Mesa+ Institute, University of Twente, 7500 AE Enschede, The Netherlands.}\affiliation{
Max Planck Institute for Dynamics and Self-Organization, 37077 G\"ottingen, Germany.}

\date{\today}

\date{\today}

\begin{abstract}
\textcolor{black}{We report on the Lagrangian statistics of acceleration of small (sub-Kolmogorov)
%~($\Xi \equiv d_p/\eta<$~1)
bubbles and tracer particles with Stokes number St $\ll$1 in turbulent flow.
 %in the range~130~--~300.
%We use a high power laser with volumetric optics to illuminate the particles. We use a single high speed camera to record our particles. 
%At high Re$_\lambda$, we find that the acceleration statistics of bubbles are similar to that of tracer particles. 
\textcolor{black}{At decreasing Reynolds number,
%~(Re$_{\lambda}$)
the bubble accelerations show deviations from that of tracer particles}, i.e. they deviate from the Heisenberg-Yaglom prediction and show a quicker decorrelation despite their small size and minute St. \textcolor{black}{Using direct numerical simulations, we show that these effects arise due the drift of these particles through the turbulent flow.
%the mean drift of these \textcolor{red}{buoyant} particles.
%In this case, the bubbles show an elevated Heisenberg-Yaglom constant and their accelerations decorrelate  even quicker than that of tracer particles.
%Thus, contrary to expectation, low Stokes number~(St $\ll$ 1) and small particle size~($d_p/\eta \leq$ 1) do not ensure tracer-like behavior for light particles in turbulence. 
We theoretically predict this gravity-driven effect for developed isotropic turbulence, with the ratio of Stokes to Froude number or equivalently the particle drift-velocity  governing the enhancement of acceleration variance and the reductions in correlation time and intermittency.} Our predictions are in good agreement with experimental and numerical results.
%The Froude number, based on the turbulent kinetic energy to the energy of rising bubble provides a good indication of this effect. 
The present findings are relevant to a range of scenarios encompassing tiny bubbles and droplets that drift through the turbulent oceans and the atmosphere. \textcolor{black}{They also question the common usage of micro-bubbles and micro-droplets as tracers in turbulence research.}}
 %is inadequate to predict the dynamics of particles in turbulence.
% indication for using have dire implications for many studies which use bubbles as tracer particles.
\end{abstract}

\maketitle

%\textit{Importance of the research $-$}
%
%\textit{Open issues $-$}
%
%
%\textit{Our approach $-$}
%
%\textit{Experiments  $-$} 
%
%\textit{Results $-$}
%
%\textit{Marginally buoyant particles - Size effect $-$}
%
%\textit{Extremely buoyant particte - Reynolds number effect $-$}
%
%\textit{Conclusions $-$}

\maketitle

\textcolor{black}{Heavy and light particles caught up in turbulent flows often behave differently from fluid tracers. The reason for this is \textcolor{black}{usually} the particle's inertia, which can drive them along trajectories that differ from those of the surrounding fluid elements~\cite{squires1991preferential,eaton1994preferential,toschi2009lagrangian,bourgoin2014focus}.} 
%Light particles in particular preferentially concentrate in high vorticity regions, which leads to an overall increase in their acceleration variance.This phenomena is well explored and is found to depend strongly on the particle's  density relative to the fluid~($\beta \equiv \frac{3 \rho_p}{\rho_f + 2 \rho_p}$) and the Stokes number. 
%Light particles ($\beta$ > 1; e.g., air bubbles in water) aggregate in high vorticity regions, while heavy particles ($\beta$ < 1; e.g., a water droplet in air) get expelled from rotating regions. Light particles therefore experience intense accelerations, a fact associated with the increased likelihood of the particle to get trapped in the fine vortical structures in the flow. 
\textcolor{black}{Due to their inertia, measured by the Stokes number St\footnote{Stokes number, $\text{St}\equiv$~$\tau_p/\tau_\eta$, where $\tau_p$ is the particle response time, and $\tau_\eta$ is the Kolmogorov time scale of the flow.},
such particles depart from fluid streamlines and distribute nonhomogeneously even when the carrier flow is statistically homogeneous~\cite{wood2005preferential, martinez2010bubble,volk2008acceleration,fiabane2012clustering,calzavarini2008quantifying,toschi2009lagrangian,lohse2008viewpoint,tanaka2008classification}). %This effect was noticeable for large St ($\geq$~1) and large $\beta$ ($>$1). 
%In these situations, light particles get entrapped in vortical structures of the flow, while heavy particles get expelled out of them resulting in accelerations very different from that of the average fluid element~\cite{toschi2009lagrangian,lohse2008viewpoint,tanaka2008classification}.}
%Higher levels of fluctuations and higher levels of intermittency are important features that distinguish light particle acceleration statistics from their heavy and neutrally buoyant counterparts~\cite{volk2008acceleration}.
\textcolor{black}{Numerical studies have captured several interesting effects of particle inertia through point-particle simulations in \textcolor{black}{homogeneous isotropic turbulence~\cite{prakash2012gravity,volk2008acceleration,cencini2006dynamics,salazar2012inertialacceleration}.} 
%As found by~\cite{calzavarini2009acceleration}, a very light particle such as an air-bubble in water experiences acceleration fluctuations $\left <a^2 \right > \sim 10$ times the tracer particle value, which is linked to the dominance of the added mass term for light particles. 
For instance, with increasing inertia, light particles showed an initial increase in acceleration variance~(up to a value $\left <a^2 \right > \sim 9$~times the tracer value) followed by a decrease, while heavy particles showed a monotonic trend of decreasing acceleration variance~\cite{calzavarini2009acceleration}.
%, due  $\beta \frac{D\bfu}{D t}$ in the particle equation of motion~$\rho_p\to0$
%However, such extreme accelerations occur only under certain conditions, which depend on the particle's size~($\Xi \equiv d_p/\eta$) and density-ratio~($\Gamma \equiv \rho_p/\rho_f$), and its Stokes number~$\text{St}\equiv$~$\tau_p/\tau_\eta$~
Such modifications of acceleration statistics arose primarily from the slow temporal response of these inertial particles, i.e when St was finite~\cite{calzavarini2008dimensionality,calzavarini2008quantifying,parishani2015effects,ireland2015effect,jung2008behavior}.}  
%While clustering has been extensively reported for light and heavy particles, its strength depends on the particle's size~($\Xi \equiv d_p/\eta$) and density-ratio~($\Gamma \equiv \rho_p/\rho_f$), and the Stokes number~St $\equiv$~$\tau_p/\tau_\eta$~
%. Here, $d_p$ is the particle diameter, $\eta$ is the Kolmogorov length scale, $\rho_p$ is the particle mass density, $\rho_f$ is the fluid mass density, $\tau_p \equiv d_p^2/(12 \beta \nu)$ is the particle response time, with $\beta \equiv 3 \rho_f/(\rho_f + 2 \rho_p)$, and $\tau_\eta$ is the Kolmogorov time scale of the flow.   
%Added mass effects are noticeable only at finite particle Stokes numbers.
% According to these, clustering is least expected for particles whose length and time scales are small compared to $\eta$ and $\tau_\eta$, respectively, of the flow. 
 %Therefore, a particle with size smaller than $\eta$ and small enough Stokes number may be expected to behave like a passive tracer regardless of its density relative to the fluid. 
%\textcolor{black}{The vast majority of studies on particles in turbulence has addressed the effects of finite inertia~(St) on particle dynamics~\cite{volk2008acceleration,bec2010turbulent,bec2007heavystokes,cencini2006dynamics,jung2008behavior,fouxon2008separation,salazar2012inertialacceleration,biferale2014intermittency}.
\textcolor{black}{In comparison, a lower limit of inertia can be imagined~(St~$\ll1$), when the particles respond to even the quickest flow fluctuations and, hence, are often deemed good trackers of the turbulent flow regardless of their density ratio~\cite{toschi2009lagrangian,calzavarini2009acceleration,calzavarini2008dimensionality, bec2006acceleration}. The widespread use of small bubbles and droplets in flow visualization and particle tracking setups~(e.g. Hydrogen bubble visualization and droplet-smoke-generators) is founded on this one assumption $-$ that St~$\ll$~1 renders a particle responsive to the \textcolor{black}{fastest fluctuations of the flow~\cite{douady1991direct,smith1983technique,lu1985image,holzner2008lagrangian,mercado2012lagrangian}.}}}

\textcolor{black}{In many practical situations, particles are subjected to body forces, typically gravitational or centrifugal~\cite{mazzitelli2004lagrangian}. This can be the case for rain droplets and aerosols settling through clouds, and tiny air bubbles and plankton drifting through the oceans~\cite{falkovich2002acceleration, falkowski2012ocean}.}
%From these studies, it was revealed that increasing the particle's inertia ($\Gamma \equiv \rho_p/\rho_f$) resulted in a monotonic decrease of $\left < a_p^2 \right >$, in agreement with the classical inertial~(temporal) filtering picture~\cite{bec2006acceleration}. \textcolor{black}{Experiments have addressed the same question by using helium-filled soap bubbles in air~\cite{qureshi2007turbulent}.
%However, these found a finite limit for $\left < a_p^2 \right > $ at large density ratios~\cite{bourgoin2011turbulent,qureshi2008acceleration}. Despite the progress in recent years, such deviations between experiments and simulations remain to be explained.}
%The most common examples of particles in turbulent flows are droplets settling through clouds
%\textcolor{black}{In such situations, the particles are subjected to body forces, typically gravitational~\cite{mazzitelli2004lagrangian}, and in some situations centrifugal~\cite{grossman2016high}. }
%This can be the case for droplets settling through clouds, and for tiny air bubbles and plankton drifting through the oceans~\cite{falkovich2002acceleration, falkowski2012ocean}.} 
\textcolor{black}{The effects of gravitational settling were first brought to light through numerical studies using random and cellular flow fields~\cite{maxey1986gravitational,maxey1987gravitational,maxey1987motion,maxey1990advection,gustavsson2014clustering}. 
%These showed preferential concentration and enhanced settling velocities for heavy particles settling in random and cellular flow fields. 
Their findings were instrumental in highlighting the role of gravity on the clustering and the enhancement of settling velocities of heavy particles in a flow. \textcolor{black}{More recently, inertial effects on settling particles were analysed using direct numerical simulations of fully developed homogeneous isotropic turbulence~\cite{bec2014gravity,good2014settling,chang2015turbulent,parishani2015effects,ireland2015effect}. These revealed that gravity can lead to major modifications of particle clustering, relative velocity and pair statistics}}, which could be characterized as a function of St and the ratio of turbulent to gravitational acceleration: $a_\eta/g$.
%, which depended on the complex interplay between turbulence~(Re$_{\lambda}$), particle inertia (St), and the ratio of turbulent to gravitational acceleration:~$a_\eta/g \equiv \epsilon^{3/4}/(g \nu^{1/4})$, where $\epsilon$ is the dissipation rate, $\nu$ is the kinematic viscosity, and $g$ is the acceleration due to gravity. 
%More recently, an approximate relation was developed, which predicted an enhancement of acceleration for heavy particles at small Stokes numbers~\cite{parishani2015effects,ireland2015effect}. While these were hypothesized to originate from the gravitational settling of particles, the predictions await experimental and numerical confirmation.} 

\textcolor{black}{While the effects of gravity on particle settling velocity  and clustering were shown to be significant~\cite{bec2014gravity,gustavsson2014clustering}, another crucial observable of Lagrangian turbulence is the particle's acceleration statistics~(variance, correlation and intermittency). 
%\textcolor{red}{How gravity affects a particle's acceleration and its higher moments in turbulent flows is not well known.} 
%{However, the effects of gravity on the acceleration statistics have received less attention, with few studies providing theoretical predictions with good experimental and numerical confirmation. 
%Obtaining the acceleration statistics can be challenging since these require accurate determination of higher derivatives, and these became possible only recently with the advent modern particle-tracking techniques~\cite{toschi2009lagrangian,brown2009acceleration,mathai2015wake}.
Acceleration is important because its variance and time correlation are tightly linked to the energy dissipation rate, a quantity central to characterizing turbulent flows. Yet another feature unique to turbulent flows is the high level of intermittency, or deviations from Gaussian statistics. Acceleration flatness is key to quantifying this for turbulent flows.} \textcolor{black}{Therefore, a generic description of these quantities for rising and settling particles of arbitrary density is desirable.}

\textcolor{black}{In this Letter, we present the Lagrangian acceleration statistics of small air-bubbles and neutrally buoyant tracer particles in a turbulent water flow where the Taylor-Reynolds number Re$_{\lambda}$ is varied in the range $130-300$. At decreasing Re$_{\lambda}$ the bubble accelerations show deviations compared to tracer particles, which occur despite their very small St~(0.004~--~0.017) and small particle size.
%Re$_\lambda$ is varied from 130 to 300. 
%Under these conditions, the particles have a small size-ratio~ ($\Xi \equiv d_p/\eta < 1$) and their Stokes number lies in the range  of 0.005$-$0.02, i.e. low enough to expect them to respond to even the smallest scale fluctuations in the flow. %The ratio St/Fr varies from 0.9 to 4.8. 
%The role of gravity in producing these deviations is explore using direct numerical simulations~(DNS), using the model eq. \ref{}.
\textcolor{black}{To explain these, we conduct DNS of particles in homogeneous isotropic turbulence, which reveals that the deviations arise due to the drift of these particles through the turbulent flow.
We develop a generic theory that predicts these gravity-induced deviations for an arbitrary density particle, with the ratio~St/Fr or equivalently the ratio of particle drift velocity to Kolmogorov velocity, as the relevant parameter controlling the deviations from ideal tracer behavior.}} \textcolor{black}{Further, we provide insight into the modification of intermittency of particle acceleration arising due to gravity.}
%We extend our theory to arbitrary density particles, which reveals that the ratio ? St/Fr can be single prominent parameter controlling the extent of deviation from ideal tracer behavior.

%We compare the acceleration variance and decorrelation times for these bubbles with those for small neutrally buoyant particles~($\Xi< 1$) under identical experiments conditions. Surprisingly, the bubbles experience higher accelerations with quicker decorrelations, despite their small $\Xi$ and St. Using direct numerical simulations, we show that these new effects are captured by the inclusion of the gravity term. We develop a theory that predicts this gravity-driven effect for all density ratio particles, with the ratio $-$ St/Fr, as the single parameter controlling the behavior. Our predictions fall in good agreement with both DNS and experimental results. Thus, we present here an interesting new direction of research on small rising/falling particles in turbulent flows.

\begin{figure} [!htbp]
	\centering
	\includegraphics[width=.66 \linewidth]{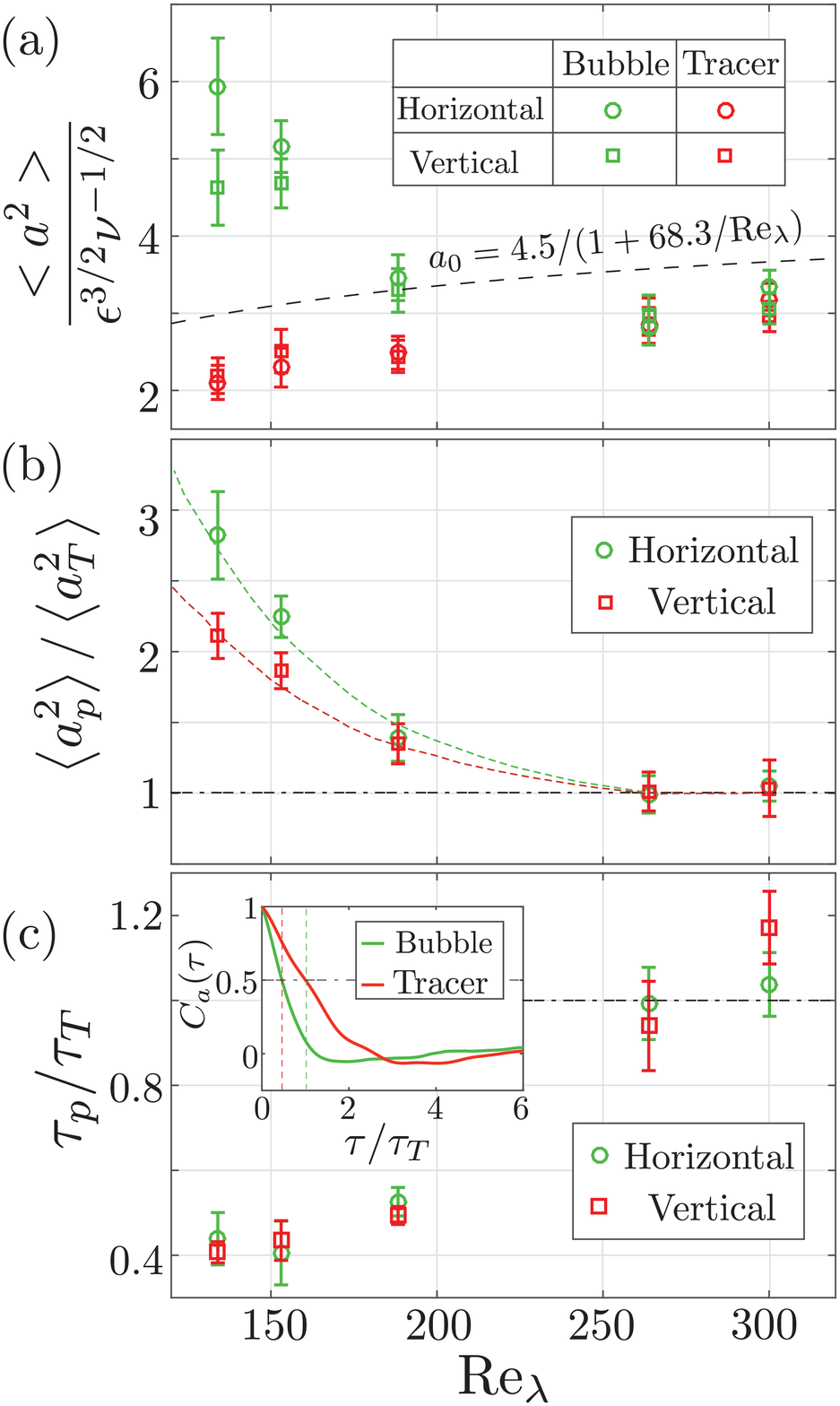}
	\caption{(a)~Heisenberg-Yaglom constant, a$_0$, estimated for bubbles and tracer particles from experiments at different Re$_{\lambda}$. The dashed curve gives the $a_0$ estimate according to~\cite{ni2012lagrangian}. (b)~Normalized acceleration variance  and (c)~correlation time of acceleration for bubbles vs Re$_{\lambda}$ from experiments. $\left <a_T^2 \right >$ is the tracer particle acceleration variance. $\tau_p$ and $\tau_T$ are defined as the 0.5~crossing time of correlation for bubbles and tracer particles, respectively. Inset: Normalized correlation function of acceleration, $C_a(\tau)$, at the lowest Re$_{\lambda} \approx 134$.}
	\label{fig:a0_a_var_vs_Re}
	%\vspace{-.5 cm}
\end{figure}

\begin{table} [!htbp]
	\caption{Flow characteristics in the Twente Water Tunnel. Re$_{\lambda}$ -- the Taylor-Reynolds number (approximate), St -- the bubble Stokes number, and $ a_\eta/g$ the ratio of turbulent to gravitational acceleration~\cite{bec2014gravity}.}
	\label{tab:FLOW} 
	\centering
	\begin{tabular}{p{0.12\linewidth}p{0.19\linewidth}p{0.14\linewidth}p{0.14\linewidth}p{0.15\linewidth}p{0.1\linewidth}}
		\hline
		\hline
		Re$_{\lambda}$ & 134 & 153 & 188 & 263 & 301\\
		%\hline St & 0.008 & 0.011 & 0.016 & 0.022  &  0.035 \\ 
		\hline St & 0.004 & 0.005 & 0.008 & 0.011  &  0.017 \\ 
		\hline $a_\eta/g$ & 0.0016 & 0.0026 & 0.0044 & 0.0072 & 0.0141 \\ 
		\hline
		$\text{St/Fr}$ & 4.899& 4.159 & 3.498 & 2.953 & 2.363 \\ 
		\hline
		\hline
	\end{tabular}
	\label{TWT_chara}
	\vspace{-.1 cm}
\end{table}

\textcolor{black}{The experiments were performed in the Twente Water Tunnel facility~(TWT), in which an active grid generated nearly homogeneous and isotropic turbulence in the measurement section~\cite{mathai2015wake}. 
%The measurement section has dimensions 2 x 0.45 x 0.45 m$^3$. An active-grid positioned upstream of the measurement section generates nearly homogeneous and isotropic turbulence in the measurement section~\cite{Mercado2012Lagrangian}.  
%Re$_{\lambda}$ was varied from 130 to 300 by either changing the grid rotation speed or by changing the mean flow speed in the water tunnel. 
The turbulent flow was characterized using hot-film anemometry technique at different Re$_{\lambda}$~(see Table~\ref{tab:FLOW}). Small bubbles~($\approx 150 \pm 25~\mu$m) were generated by blowing pressurized air through a porous ceramic plate. 
%In a separate set of experiments, neutrally buoyant tracer particles were dispersed in the flow.
The particles were imaged using a high-speed camera~(Photron PCI-1024) at a recording rate of 1000~fps. The camera moved on a traverse system, and illumination was provided using a 100 W Pulsed Laser~(Litron~LDY-303HE)~(see Fig.~1 in supplemental material~\cite{supplemental}).
%Particle tracking was performed as the camera moved at constant velocity matching the mean flow speed in the measurement section~\cite{prakash2012gravity}.
%\textcolor{red}{The traverse could move vertically a distance of about 1.5~m, and its motion consisted of an accelerating, a constant-velocity and a decelerating phase. Particle tracking was performed during the constant velocity phase of the traverse motion.} 
Similar moving camera setups have been used with heavy particles in a wind tunnel~\cite{ayyalasomayajula2006lagrangian}, however, the tracking duration was short in those experiments. In the present case, we introduce a novel experimental modification, which allows for long particle trajectories to be recorded. We placed a mirror inside the tunnel at 45$^\circ$ inclination with the horizontal~(see Fig.~1 in supplemental material~\cite{supplemental}). The laser beam was expanded into a volume, which was reflected vertically by the mirror. 
%This produced a vertical volume illumination of nearly uniform intensity and  enabled us to track the particles for several~$\tau_\eta$. %Recording was stopped when the camera moved closer than 0.4~m from mirror. At this distance, disturbances due to the downstream mirror were expected to be minor. 
\textcolor{black}{Obtaining the acceleration from position data requires accurate determination of higher derivatives.  We have here combined the conventional Gaussian-kernel smoothening method with a smoothing-spline based technique~\cite{brown2009acceleration,mathai2016translational} which eliminated biases due to a~priori choice of filter windows and ensured reliable estimates of the acceleration}.} 
%Experiments were also conducted using fluorescent neutrally buoyant tracer particles under identical experimental conditions.

%\textcolor{red}{The flow characterization was done with the mirror in place and the flow characteristics did not vary much in the region of particle tracking.}

%Lagrangian particle tracking experiments are carried out using a high-speed camera (Photron PCI-1024). The recording rate was about 1000 frames per second (fps) at 1 Megapixel resolution. This framing rate ensured at least 20 images per Kolmogorov time. This ensured that even the fastest fluctuations were also captured. The camera is focused on a 1-2 cm thick plane in the middle of the measurement section and the recording window is about 30~x~30~mm$^2$ area, generating a spatial resolution of 30~$\mu$m/pixel. The aperture is set so that particles are sharply in focus within this measurement plane.  Illumination is provided using a 100~W Pulsed Laser~(Litron~LDY-303HE), which is synchronized with the camera. The error in determining the particle's position is sub-pixel. On average the trajectories had a length of 700 data points (700~ms.) Any trajectory that has a length below 100 data-points is discarded. This conservative stand ensures that our trajectories have a duration of at least 1.5$\tau_\eta$, allowing more reliable determination of higher order derivatives. The trajectories are smoothed using a Gaussian-Kernel smoothing~\cite{Voth2002measurement}.

\textcolor{black}{We first address the question of how the bubble accelerations compare to that of similar-sized neutrally buoyant particles at different Re$_{\lambda}$. According to the prediction by Heisenberg and Yaglom~\cite{la2001fluid}, the \textcolor{black}{single-component variance of acceleration} should follow the relation $\left < a^2 \right > = a_0 \epsilon^{3/2} \nu^{-1/2}$, where $\epsilon$ is the dissipation rate, and $\nu$ the kinematic viscosity. In Fig~\ref{fig:a0_a_var_vs_Re}(a), we plot the Heisenberg-Yaglom constant $a_0$ for bubbles along with that for tracer particles. At high Re$_{\lambda}$, the bubbles behave similarly to tracers, with comparable $a_0$. However, at low Re$_\lambda$ the bubbles show deviations from tracers, with an elevated $a_0$. The horizontal acceleration shows the greatest deviation, with $a_0 \approx 6$, while for the vertical component, $a_0 \approx 4.5$ at Re$_{\lambda} \approx 134$. For neutrally buoyant tracer particles, $a_0$ is lower, $\approx$~2.1 at Re$_{\lambda} \approx 134$, and shows a marginal increase with Re$_{\lambda}$. Thus, the horizontal component of the acceleration variance for bubbles is almost three times that of the tracer value at the lowest Re$_{\lambda}$~(see Fig.~\ref{fig:a0_a_var_vs_Re}(b)).}
% since we normalized with the tracer particle value in the corresponding directions.
\textcolor{black}{This is also reflected in the correlation time for bubbles~(Fig.~\ref{fig:a0_a_var_vs_Re}(c) \& Inset), which is shorter compared to that of tracers. \textcolor{black}{These deviations are surprising, since the lowest Re$_\lambda$ corresponds also to the smallest St in our experiments~(see Table.~\ref{TWT_chara}).}}
% the particle Stokes numbers are the smallest at low Re$_{\lambda}$ (see Table.~\ref{TWT_chara}). We note that 
%added mass effects that could enhance the acceleration variance and decrease the correlation time of bubbles, are least expected at these minute Stokes numbers~(see Table.~\ref{TWT_chara}).}

%
%We first address the question of how bubble accelerations compare to that of neutrally buoyant tracer particles at different Re$_{\lambda}$. In Fig.~\ref{fig:a0_a_var_vs_Re}(a), we show the Heisenberg-Yaglom constant prediction for bubbles and tracers. For bubbles, we observe two major deviations from tracer particles. First, the a$_0$ for bubbles always exceeds that for tracer particle. The deviation is greatest at lower Re$_{\lambda}$, with a$_0 \approx$ 6 as compared to a$_0 \approx$ 2.1 for the tracer particle. Second, we find that a$_0$ in horizontal direct exceeds a$_0$ in vertical direction. In Fig.~\ref{fig:a0_a_var_vs_Re}(b), we plot the tracer-normalized acceleration variance for the bubbles. Bubble acceleration variance reaches values as high as 3 times the tracer value at lowe Re$_{\lambda}$. The tracer-normalized decorrelation time, $\tau_p/\tau_T$ declines below 1 at low Re$_{\lambda}$. Surprisingly, these deviations are most pronounced at low Re$_{\lambda}$, when the bubble St and $\Xi$ are the extremely small.

\begin{figure} [!htbp]
\centering
\includegraphics[width=.67 \linewidth]{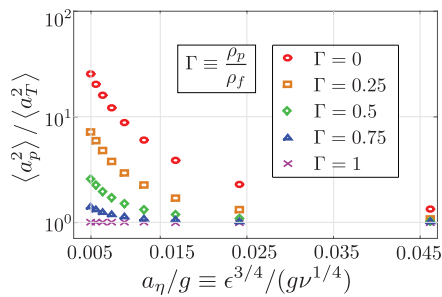}
\caption{Normalized horizontal acceleration variance for buoyant particles vs  $a_\eta/g$ obtained from Eulerian-Lagrangian DNS at Re$_{\lambda} \approx 80$.}
\label{fig:a_var_vs_Re_DNS}
\end{figure}

From Fig~\ref{fig:a0_a_var_vs_Re}(b), we also note that the vertical component of acceleration is consistently lower as compared to the horizontal one. \textcolor{black}{This anisotropy is not inherent in the carrier flow~\cite{poorte2002experiments} and therefore suggests the role of gravity~\cite{maxey1987gravitational}.} We note that with decreasing Re$_{\lambda}$, the ratio of turbulent to gravitational acceleration, $a_\eta/g$, decreases in our experiments~(Table.~\ref{TWT_chara}). In order to investigate this effect in a systematic way, we perform DNS of homogeneous isotropic turbulence at Re$_{\lambda} \approx 80$ in the presence of gravity. For the particles, we use a model considering a dilute suspension of passively advected point-spheres acted upon by inertial and viscous~(Stokes drag) forces~(see supplemental material~\cite{supplemental}). We neglect lift, history and finite-size Fax\'en forces, since these are verified to be negligible in the point particle limit~\cite{calzavarini2009acceleration,calzavarini2012impact}. The model equation of motion for a small inertial particle advected by a fluid flow field,~($\bfU(\bfX(\mathcal{T}),\mathcal{T})$), may be written as:
\begin{equation}
\ddot{\bfX} = \frac{3 \rho_f}{\rho_f + 2 \rho_p}\left( \frac{D\bfU}{D \mathcal{T}}  + \frac{12 \nu}{d_p^2}(\bfU - \dot{\bfX}) +  g\ \hat{\bfe}_{z} \right) - g\ \hat{\bfe}_{z},
\label{model_equation}
\end{equation} 
where $\rho_f$ and $\rho_p$ are the fluid and particle mass densities, respectively, $d_p$ is the particle diameter, and $\hat{\bfe}_{z}$ is the unit vector in the direction of gravity. \textcolor{black}{The particles under consideration are buoyant~($0 \leq \Gamma < 1 $) and have very small \textcolor{black}{Stokes number~($\text{St} \approx 0.05$).}} We vary the gravity intensity $g$ for these particles, resulting in a range of values for $a_\eta/g$, according to the Froude number definition in~\cite{bec2014gravity,ireland2015effect}. We first address the case of bubbles~($\Gamma = 0$ in Fig.~\ref{fig:a_var_vs_Re_DNS}) at various strengths of $g$. With increasing $g$, we recover the trends observed in our experiments, i.e. the bubble acceleration variance increases.
%, and they show a decreased correlation time~(Inset to Fig.~\ref{fig:a_var_vs_Re_DNS}). 
Gravity enhances the acceleration in both vertical and horizontal directions, and this is accompanied by a decrease in correlation time~(see supplemental material~\cite{supplemental}). 
%These results are in qualitative agreement with our experiments~(Fig.~\ref{fig:a0_a_var_vs_Re}(b) \& (c)).

Fig.~\ref{fig:a_var_vs_Re_DNS} clearly demonstrates the role of gravity in enhancing the particle's acceleration variance. \textcolor{black}{At the same time, we note that the degree of enhancement diminishes with growing $\Gamma$ even at fixed $a_\eta/g$. Also, the effect of gravity on ${\left <a_p^2 \right >}/{\left < a_T^2 \right >}$ appears more pronounced in our simulations~(see Fig.~\ref{fig:a0_a_var_vs_Re}(b)) and Table~\ref{TWT_chara}).} Thus, the Froude number, if defined as $a_\eta/g$~\cite{bec2014gravity,ireland2015effect}, can only give a qualitative prediction of the gravity effect. An exact prediction for an arbitrary-density particle is missing.

For a better appreciation of the contribution of gravity to the particle dynamics, we non-dimensionalize eq.~\eqref{model_equation} in terms of the Kolmogorov length~$\eta$ and time scales~$\tau_\eta$. We obtain
%\begin{equation}
%\ddot{\bfx} = \beta \frac{D\bfu}{D t}  + \frac{1}{\tau_p}(\bfu - \dot{\bfx}) + (\beta - 1) g \hat{\bfz}
%\end{equation}
%In dimensionless units
\begin{equation}
\ddot{\bfx} = \beta \frac{D\bfu}{D t}   + \frac{1}{\text{St}}(\bfu -  \dot{\bfx}) + \frac{1}{\text{Fr}} \hat{\bfe}_{z},
\label{eq_non-dim}
\end{equation}
%, with an additional ($\beta-1$) in the denominator.
%\begin{equation}
%\beta \equiv  \frac{3 \rho_f}{\rho_f + 2 \rho_p}; \quad \text{St} \equiv  \frac{3 \beta \nu}{a^2 \tau_{\eta}} = \frac{\tau_p }{ \tau_{\eta}} ; \quad \text{Fr} \equiv  \frac{ a_{\eta}  }{ \left( \beta - 1 \right) g }
%\end{equation}
\textcolor{black}{where St $\equiv d_p^2/(12 \beta \nu \tau_\eta)$ is the Stokes number and $\text{Fr} \equiv a_\eta/((\beta-1)g)$ is a buoyancy-corrected Froude number that takes the particle density, through $\beta~\equiv~3\rho_f/(\rho_f~+2~\rho_p)$, into account.}
\textcolor{black}{In this situation, two important small-Stokes limits may be considered~\cite{maxey1987gravitational,parishani2015effects}}.
At high turbulence intensities (Fr~$\to \infty$), the third term on the right-hand-side of eq.~(\ref{eq_non-dim}) may be neglected. 
This leads to  the well known result $\ddot{\bfx}\simeq D_t \bfu$ for particle acceleration, where  $D_t \bfu$  is the fluid tracer acceleration.
%along a fluid streamline.\\
However, for small Fr, the small St limit leads to the result~$\ddot{\bfx} = D_t \bfu~+~\frac{\text{St}}{\text{Fr}} \partial_z \bfu$ for particle acceleration. By employing results from homogeneous isotropic turbulence (see supplemental material~\cite{supplemental}), one obtains the following relations linking the acceleration variance of particles to that of the fluid tracer:

\begin{eqnarray}
\frac{\left <a_h^2 \right >}{\left < a_T^2 \right >} \equiv \frac{ \langle \ddot{x}^2  \rangle}{ \langle (D_t u_x)^2  \rangle} &\simeq& 1 +  \frac{2}{15 a_0} \left( \frac{\text{St}}{\text{Fr}}\right)^2,\\
\frac{\left <a_v^2 \right >}{\left < a_T^2 \right >} \equiv \frac{ \langle \ddot{z}^2  \rangle}{ \langle (D_t u_x)^2  \rangle} &\simeq& 1 +  \frac{1}{15 a_0} \left( \frac{\text{St}}{\text{Fr}}\right)^2,
\label{variance_theory_1}
\end{eqnarray}
\textcolor{black}{where $a_h$ and $a_v$ are the horizontal and vertical accelerations, respectively, for an arbitrary-density particle, $a_T$ is the tracer particle acceleration, and $x$ and $z$ represent the horizontal and vertical directions, respectively.} %$a_0$ is the Heisenberg-Yaglom constant, which has a weak dependence on Re$_{\lambda}$, but is practically constant at very large Re$_{\lambda}$~\cite{la2001fluid,voth2002measurement}. }
	
%A similar functional dependence was derived for heavy particles based on a terminal velocity argument in~\cite{parishani2015effects}. However, these predictions required a measure of the terminal velocity and the dissipation rate $\epsilon$ of the turbulent flow.}

\begin{figure} [!htbp]
	\centering
	\includegraphics[width=0.7 \linewidth]{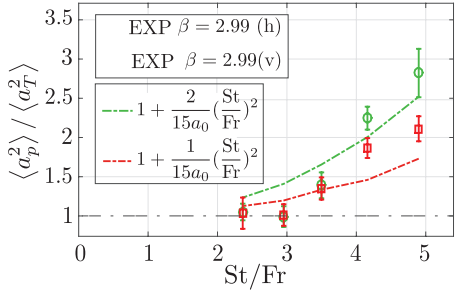}
	\caption{Normalized acceleration variance in horizontal (h) and vertical (v) directions vs St/Fr for bubbles from our experiments. The dashed lines give the predictions based on the eq.~(3)~\&~(\ref{variance_theory_1}),  using $a_0$ obtained for the tracer particles in the present experiments~(Fig.~\ref{fig:a0_a_var_vs_Re}(a)).}
	\label{fig:EXP_Theory}
	\vspace{0 cm}
\end{figure}

In Fig.~\ref{fig:EXP_Theory}, we compare the normalized acceleration variance vs St/Fr from experiment with our theoretical predictions~(eq.~(3)~\& (\ref{variance_theory_1})). 
%The solid lines show the quadratic dependence on St/Fr, with $a_0 =2.14$ obtained from DNS at Re$_{\lambda} \approx 80$. This value of $a_0$ over-predicts the acceleration variance, particularly for the larger St/Fr values. We note, however, that the flow conditions and Re$_{\lambda}$ are different in our experiments. 
The dashed lines show the theoretical predictions using the $a_0$ from the present experiments~(Fig.~\ref{fig:a0_a_var_vs_Re}(a)). The experimental data-points are in reasonable agreement with our predictions. 
%\textcolor{black}{Therefore, one needs to consider the ratio St/Fr, as opposed to St and Fr separately, in order to gauge these gravity-driven effects properly.
\textcolor{black}{Therefore, the apparent Re$_{\lambda}$ dependence that was seen in our water tunnel experiments~(Fig.~\ref{fig:a0_a_var_vs_Re}) is in fact a St/Fr effect, since the St/Fr increases with decreasing Re$_{\lambda}$ in our experiments~(see Table~\ref{tab:FLOW}).}

%The deviations are possibly due to errors in estimation of St/Fr and the neglected lift force~$\frac{\beta}{3}(u -\dot{\bfx}) \times \omega$~\cite{mazzitelli2003relevance}, which has a contribution to the acceleration~(see supplemental~\cite{supplemental}).

\begin{figure} [!htbp]
\centering
\includegraphics[width=.95\linewidth]{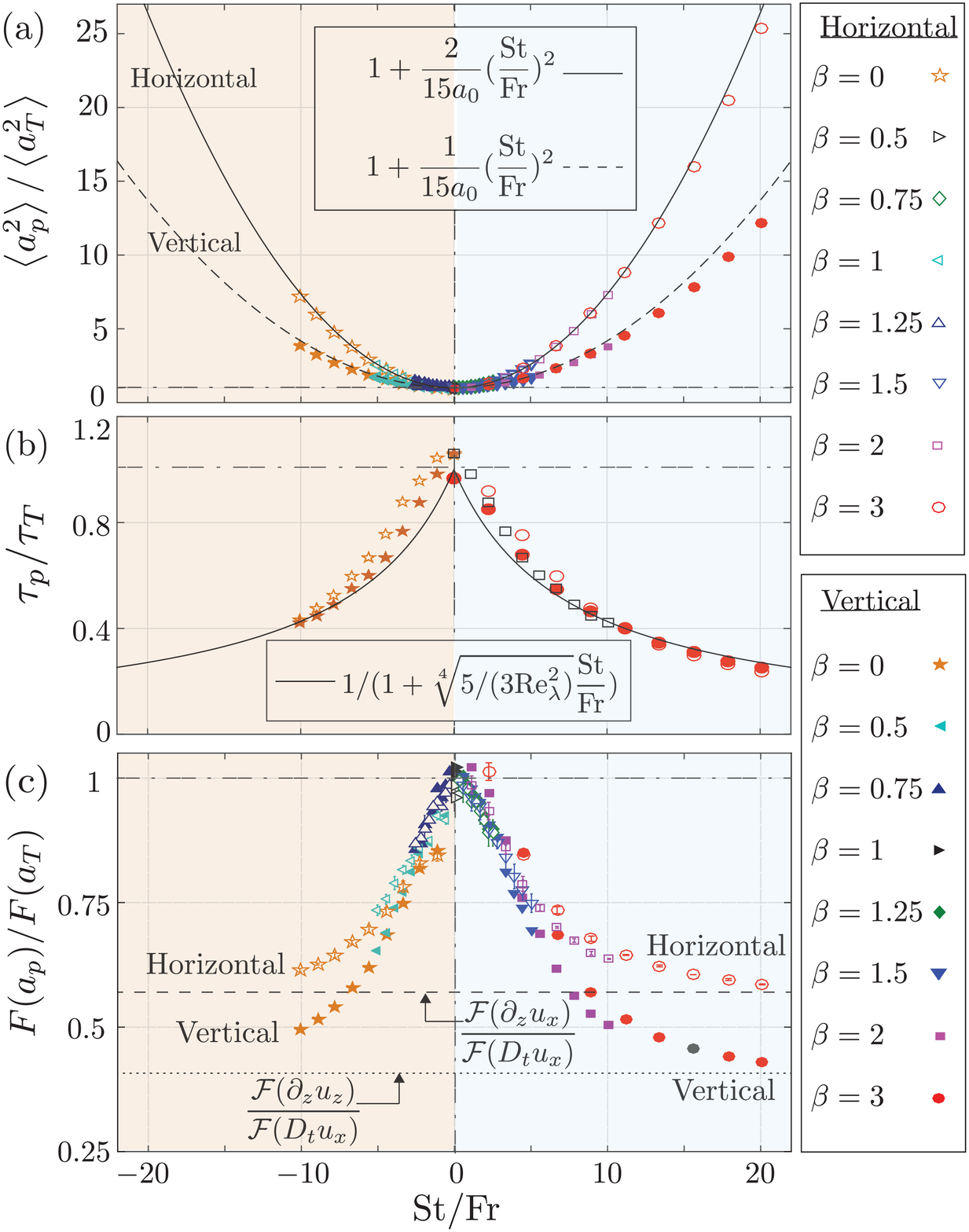}
\caption{(a) Normalized acceleration variance, (b) normalized correlation time, and (c) normalized flatness factor vs St/Fr for a family of buoyant and heavy particles, obtained from DNS. Solid and dashed curves in (a) show our theory-predictions for horizontal and vertical accelerations, respectively~(eq.~(3)~\&~(\ref{variance_theory_1})). (b) \textcolor{black}{The black curve shows the theoretical prediction for correlation-time. (a)-(c) Hollow and solid symbols correspond to horizontal and vertical components, respectively.}}
\label{fig:variance_numerics}
%\vspace{-.5cm}
\end{figure}

The present bubble-tracking experiments cover a narrow range of $\text{St/Fr}=[2-5]$ at fixed density-ratio~($\beta=2.99$), while  eq.~(3)~\& (\ref{variance_theory_1}) should be valid for arbitrary density-ratio. \textcolor{black}{To test this, we compare the results of DNS for an extended range of density-ratios~$\beta = [0, 3]$ and $\text{St/Fr} = [-10, 20]$.} In Fig.~\ref{fig:variance_numerics}(a),
%In Fig.~\ref{fig:variance_numerics}(a) we plot the normalized acceleration variance against St/Fr for light ($1<\beta <3$) and heavy ($0<\beta<1$) particles. 
the left half~($\text{St/Fr} <0$) points to heavy particles, and the right one~($\text{St/Fr} >0$), to buoyant particles.
%The numerical simulations are in excellent agreement with the theoretical predictions for all the density ratios studied. 
\textcolor{black}{The predicted quadratic dependence on St/Fr and even the pre-factors $2/(15a_0)$ and $1/(15a_0)$ for the horizontal and vertical components, respectively, are in \textcolor{black}{agreement} with our simulations.} \textcolor{black}{We note that for sufficiently large Reynolds numbers, $a_0$ is practically constant~\cite{la2001fluid}, and the ratio St/Fr becomes the sole control parameter governing the enhancement of acceleration variance. Therefore, our results have broad applicability, to even large Re$_{\lambda}$ atmospheric and oceanic flows. In Fig.~\ref{variance_theory_1}(b), we present the numerical results for the tracer-normalized correlation time in the presence of gravity. We propose a model for the correlation time based on the time a particle takes to cross an energetic eddy of the flow~(see supplemental material~\cite{supplemental}). The model predicts a behavior of the form $\tau_p/\tau_T \approx \frac{1}{1 +  \text{k (St/Fr})}$, where $\text{k} \approx \sqrt[4]{\frac{5}{3 \text{Re}_{\lambda}^2}}$. The numerical results are in reasonable agreement with our predictions. Small deviations are noticeable in the small St/Fr range. In this range, the fluid acceleration $D_t \bf u$ dominates over the the velocity gradient $\frac{\text{St}}{\text{Fr}} \partial_z \bf u$. This explains the slower decline than what is predicted by our model in the small St/Fr range~(see supplemental material~\cite{supplemental}). We also notice that the horizontal component (hollow symbols) is slightly higher that the vertical one (solid symbols) in the small St/Fr range. This is because the transverse velocity gradients $\partial_z u_x$ are longer correlated than the longitudinal velocity gradients $\partial_z u_z$~(see supplemental material~\cite{supplemental}).}
	
%	The model predicts two regimes: a linear decline of the form $\tau_p/\tau_T \approx 1- \text{k (St/Fr})$ for moderate St/Fr, followed by $\tau_p/\tau_T \approx \frac{1}{ \text{k (St/Fr})}$ in the large St/Fr limit, where $\text{k} \approx \sqrt[4]{\frac{5}{3 \text{Re}_{\lambda}^2}}$~(see inset to Fig.~\ref{variance_theory_1}(b)).}

%For light particles, we witness two interesting competing effects for heavy particles. Whether the particles acceleration in enhance particle acceleration for a light particle under gravity. At increasing St, the added mass term increases the particle's acceleration variance, while it leads to slower repsonse to the drifting eddies.}

%\textcolor{black}{On the experimental side, we have confirmed the enhancement of acceleration variance using tiny air-bubbles dispersed in our water tunnel facility. On the heavy-particle front (left-branch of Fig.~\ref{fig:variance_numerics}(a) \& (b)), our predictions remain to be experimentally verified.}

% We note that for sufficiently large Re$_{\lambda}$, as in atmospheric and oceanic turbulence, the ratio St/Fr becomes the sole control parameter governing the normalized acceleration variance~($a_0$ may be constant for high enough Re$_{\lambda}$~\cite{la2001fluid,voth2002measurement}). \textcolor{black}{Therefore, it is of great importance to consider the effect of gravity on particle dynamics in such flows.}

While the effects of gravity on acceleration variance and correlation time have been comprehensively demonstrated, its role on the intermittency of particle acceleration is not clear. Intermittency, i.e. the observed strong deviations from Gaussianity, can be characterized in terms of the flatness of acceleration~$\mathcal{F}(a_p) \equiv \left <a_p^4 \right >/\left < a_p^2 \right >^2$. \textcolor{black}{ Assuming statistical independence between $D_t u_i$ and $\partial_z u_i$, we obtain the tracer-normalized flatness of the particle acceleration, $\mathcal{F}(a_p)/ \mathcal{F}(a_T)$, as a decreasing function of St/Fr~(see supplemental material~\cite{supplemental}). At large \text{St/Fr}, we asymptotically approach the limits}

\begin{eqnarray}
\frac{ \mathcal{F}( a_h)}{   \mathcal{F}(a_T) } \equiv \frac{ \mathcal{F}( \ddot{x})}{   \mathcal{F}(D_t u_x) } &\simeq&  \frac{ \mathcal{F}(\partial_z u_x)}{   \mathcal{F}(D_t u_x) },\\
\frac{ \mathcal{F}( a_v)}{   \mathcal{F}(a_T) } \equiv \frac{ \mathcal{F}( \ddot{z})}{   \mathcal{F}(D_t u_x) } &\simeq&  \frac{ \mathcal{F}(\partial_z u_z)}{   \mathcal{F}(D_t u_x) },
\label{flatness_theory_largeStbyFr}
\end{eqnarray}
%It is well known that $ \mathcal{F}(D_t u_x)  >  \mathcal{F}(\partial_z u_x) >\mathcal{F}(\partial_z u_z) $~\cite{ishiharaJFM2007}, meaning that particle rise/fall leads to a reduction in  flatness factor, which asymptotically approaches eq.~\ref{flatness_theory_largeStbyFr}. It is also verified that $\mathcal{F}(\partial_z u_x) >  \mathcal{F}(\partial_z u_z) $, which leads to the result $\mathcal{F}( \ddot{x}) > \mathcal{F}( \ddot{z})$. 
\textcolor{black}{It is verified that $ \mathcal{F}(\partial_z u_z) < \mathcal{F}(\partial_z u_x) < \mathcal{F}(D_t u_x)$~\cite{ishihara2007small}. This leads to the prediction $\mathcal{F}( a_v) < \mathcal{F}( a_h)$, i.e. vertical acceleration is less intermittent as compared to the horizontal one.} 
%These predictions are confirmed by our simulations, as revealed in Fig.~\ref{fig:variance_numerics}(b) for an extended $\beta-\text{St/Fr}$ range. A quantitative agreement is missing, since errors are likely in the estimation of higher moments. Nevertheless, these results showcase the first evidence of intermittency reduction even for tiny buoyant particles in turbulent flows. 
\textcolor{black}{In Fig.~\ref{fig:variance_numerics}(c), we present the normalized flatness factor from our simulations for an extended ($\beta, \text{St/Fr}$) range.  $ \mathcal{F}(a_p) $ decreases  for both buoyant and heavy particles, and the curves asymptotically reach the limits suggested by eq.~(5)~\&~(\ref{flatness_theory_largeStbyFr}). The present findings showcase the first evidence of intermittency reduction even for small St particles in turbulence.} 

\textcolor{black}{Our results show that acceleration statistics~(variance, correlation, and intermittency) is very sensitive to the ratio St/Fr. To explain the origin of these in physical terms, we consider the case of a particle drifting through a turbulent flow. As the particle drifts through the flow, it meets different eddies. Owing to its short response time, the particle readjusts to the velocity of these eddies. The rate at which the particle readjusts to the new eddies is linked to the spatial velocity gradients of the turbulent flow. As a consequence, the particle experiences accelerations that the regular fluid element does not experience, thereby increasing its fluctuations~(variance). The effect becomes prominent when the drifting time of the particle past the most energetic eddies (Taylor micro-scale eddies) becomes shorter than the timescale of these eddies of the flow, or when $\text{St/Fr} >1$. This explains the scaling of the decorrelation time in Fig.~\ref{variance_theory_1}(b)~(see supplemental material~\cite{supplemental} for details).}	

\textcolor{black}{The same physical mechanisms could explain the decline in intermittency of particle acceleration. A drifting particle, instead of probing the accelerations of fluid elements, begins to sample the spatial gradients of the flow. An interesting analogy may be drawn to the intermittency of acceleration recorded by a hot-wire probe placed in a high-speed wind/water tunnel flow, where the probe effectively registers only the spatial gradients~\cite{sandborn1976effect}. For a turbulent flow, the intermittency of the spatial gradients of velocity is lower as compared to the intermittency of the fluid element acceleration~\cite{ishihara2009study,toschi2009lagrangian}. Hence the observed decline in intermittency, which asymptotically approaches the expressions given by eq. (5) and (6) in the limit of large St/Fr. This gravity effect will be important even for moderate Stokes number particles, and the same qualitative trends may be expected. However, a moderate St particle responds slower to the turbulent eddies it drifts through. Hence, we expect the gravity effect to be less prominent than that for the $\text{St} \ll 1$ particles we presented here, and this will be the focus of a future investigation.}

\textcolor{black}{In summary, the acceleration statistics of small Stokes number particles in turbulence is greatly modified in the presence of gravity. We report three major effects: an increase in acceleration variance, a decrease in correlation time, and a reduction of intermittency for buoyant and heavy particles. The ratio $\text{St}/\text{Fr}$ governs the extent of this modification, as confirmed by our experiments using tiny air bubbles in water.} Our theoretical predictions have broad validity -- to particles of arbitrary density and even at large Reynolds numbers.
%\textcolor{black}{On the heavy-particle front, our predictions are awaiting experimental confirmation.}
Thus, a tiny bubble or droplet is not necessarily a good tracer of turbulent acceleration.
This can be important for bubbles and droplets that drift through the turbulent oceans~($g/a_\eta  \approx 100-1000$) and clouds~($g/a_\eta  \approx~10-100$)~\cite{devenish2012droplet}.
%We estimate that $g/a_\eta  \approx$ 500 in the oceans, and it lies in the range 20 --100 for atmospheric clouds~\cite{Ireland_collinsarXiv2015}. Clearly, this has relevance in scenarios where tiny bubbles and rain droplets drift through the turbulent oceans and the atmosphere.  
On the practical side, our findings point to an important  \textcolor{black}{consideration} when choosing bubbles or droplets for flow visualization and particle tracking in turbulent flows~\cite{douady1991direct}.

\textcolor{black}{We gratefully acknowledge A.~Prosperetti, S.~S.~Ray, G.~D.~Jin, S.~Wildeman, and S.~Maheshwari for insightful discussions. 
We thank S.~Huisman, X.~Zhu, P. Shukla and V.~N.~Prakash for comments which helped improve our manuscript. We thank the two anonymous referees for comments which helped improve our manuscript.}
%We thank G.-W. Bruggert, M.~Bos and B.~Benschop for technical support. 
%This work was financially supported by the Simon Stevin Prize of the Technology Foundation STW of The Netherlands, European High-Performance Infrastructures in Turbulence~(EUHIT), and COST action MP1305.

\bibliography{mybib_new_f}

\end{document}

%\title{Gravity-Driven Accelerations of Small Buoyant Particles in Turbulence}
%\title{Gravity-Driven Accelerations of Small Particles in Turbulence}
%\title{Tiny Particles do not Always Trace the Turbulent Flow}
%\title{Why Buoyancy is Prohibitive when Using Microbubbles and Microparticles \\ as Tracers in Turbulence}
\title{Micro-bubbles and Micro-particles are Not Faithful Tracers of Turbulent Acceleration\\
\large{(Supplementary material)}}

\author{Varghese Mathai}
\affiliation{Physics of Fluids Group, Faculty of Science and Technology,
Mesa+ Institute, University of Twente, 7500 AE Enschede, The Netherlands.}

\author{Enrico Calzavarini}
\thanks{enrico.calzavarini@polytech-lille.fr}
\affiliation{ 
Univ. Lille, CNRS, FRE 3723, LML, Laboratoire de Mecanique de Lille, F 59000 Lille, France}
%Laboratoire de M\'ecanique de Lille CNRS/UMR 8107, Universit\'e Lille 1 et \'Ecole Polytechnique Universitaire de Lille, Cit\'e Scientifique Avenue Paul Langevin, 59650 Villeneuve d'Ascq, France.}

\author{Jon Brons}
\affiliation{Physics of Fluids Group, Faculty of Science and Technology,
Mesa+ Institute, University of Twente, 7500 AE Enschede, The Netherlands.}
\affiliation{ Department of Mathematics and Physics, Faculty of Engineering and Computing, Coventry University, United Kingdom}

\author{Chao Sun}
\thanks{chaosun@tsinghua.edu.cn}
\affiliation{Center for Combustion Energy and Department of Thermal Engineering, 
Tsinghua University, 100084 Beijing, China.}
\affiliation{Physics of Fluids Group, Faculty of Science and Technology,
Mesa+ Institute, University of Twente, 7500 AE Enschede, The Netherlands.}

\author{Detlef Lohse}
\affiliation{Physics of Fluids Group, Faculty of Science and Technology,
Mesa+ Institute, University of Twente, 7500 AE Enschede, The Netherlands.}
\affiliation{
Max Planck Institute for Dynamics and Self-Organization, 37077 G\"ottingen, Germany.}

\date{\today}

\maketitle

\section{Experimental setup}

\begin{figure} [!htbp]
	\centering
	\includegraphics[width=0.85\linewidth]{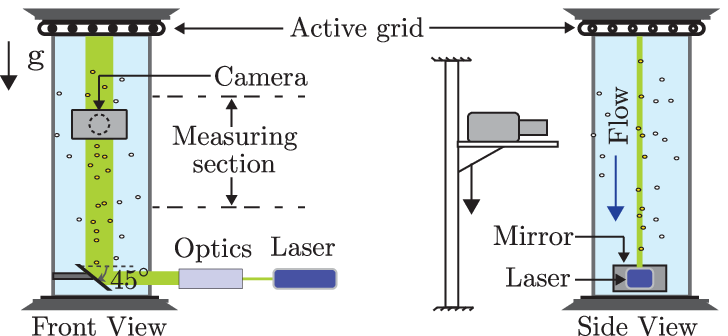}
	\caption{Schematic of the measurement section of the Twente Water Tunnel. 
		%Front and side views of the Twente Water Tunnel measurement section are shown here. 
		Bubbles and neutrally buoyant tracer particles of diameters $\approx~150~\pm$~25~$\mu$m and $\approx 125$~ $\mu$m, respectively, were dispersed in the flow for particle tracking experiments.
		%Tiny bubbles~($D\approx 150 \pm 25 \mu$m) are generated from porous plates using pressurized air. The bubbles are illuminated using an arrangement consisting of a high-speed laser and volumetric optics. The camera is mounted on an arm which moves at constant velocity matching the flow speed. This ensures long particles trajectories. The particle size in the schematic has been greatly exaggerated for ease of comprehension. 
	}
	\label{fig:SETUP}
\end{figure}

\section{Particle Equation of Motion}
The model equation of motion for a small  inertial spherical particle advected by a fluid flow field, with velocity $\bfU(\bfX(\mathcal{T}),\mathcal{T})$, is:
\begin{equation}
\mathcal{V}\ \rho_p\ \ddot{\bfX} =   \mathcal{V}\ \rho_f \frac{D\bfU}{D \mathcal{T}}  + \bf{F}_{AM}  + \bf{F}_{D} + \bf{F}_{B}  
\end{equation}
where  $\mathcal{V} = \frac{4}{3} \pi a^3$ is the particle volume, with $a$ being the particle radius, and $\rho_f$ and $\rho_p$ the fluid and particle mass densities, respectively. The forces contributing on the right-hand-side besides the fluid acceleration (which includes the pressure gradient term) are the added mass $\bf{F}_{AM}$, the drag force $\bf{F}_{D}$, and the buoyancy $\bf{F}_{B}$~\cite{maxey1983equation,gatignol1983faxen,spelt1997motion}:
\begin{eqnarray}
%\bf{F}_{FA}  &=&  \mathcal{V}\ \rho_f \frac{D\bfU}{D \mathcal{T}}\\
\bf{F}_{AM} &=&  \mathcal{V}\ \rho_f C_M \left( \frac{D\bfU}{D \mathcal{T}} -  \ddot{\bfX} \right),\\
\bf{F}_{D} &=& 6\ \pi\ \mu\ a\ (\bfU - \dot{\bfX}),\\
\bf{F}_{B} &=&  \mathcal{V}\ (\rho_p - \rho_f)\ g\ \hat{\bfe}_{z},
\end{eqnarray}
where $\mu$ is the dynamic viscosity, $g$ is the gravity intensity and $\hat{\bfe}_{z}$ is the unit-vector in the direction of gravity. Note that we use the inviscid added mass coefficient for a sphere, i.e $C_M=1/2$.
This leads to
%\begin{equation}
%\rho_p\ \ddot{\bfX} =  \rho_f\left( \frac{D\bfU}{D \mathcal{T}}  +  C_M ( \frac{D\bfU}{D \mathcal{T}}  - \ddot{\bfX})\right) +  \frac{3 \mu}{a^2}(\bfU - \dot{\bfX})
% + \left(\rho_p - \rho_f \right) g\ \hat{\bfe}_{z} 
%\end{equation}

\begin{equation}
\ddot{\bfX} = \frac{3 \rho_f}{\rho_f + 2 \rho_p}\left( \frac{D\bfU}{D \mathcal{T}}  + \frac{12 \nu}{d_p^2}(\bfU - \dot{\bfX}) +  g\ \hat{\bfe}_{z} \right) - g\ \hat{\bfe}_{z},
\label{model_equation}
\end{equation}
where $\nu \equiv \mu/\rho_f$ is the kinematic viscosity, and $d_p$ is the particle diameter. Here we neglect lift, history, and finite-size Fax\'en forces, since these are verified to be small in point-particle limit and when the particle size is smaller than the Kolmogorov length scale $\eta$ of the flow~\cite{calzavarini2012impact,calzavarini2009acceleration,tagawa2012three}.

\section{Froude number effect}

\begin{figure} [!htbp]
\centering
\includegraphics[width=0.85\linewidth]{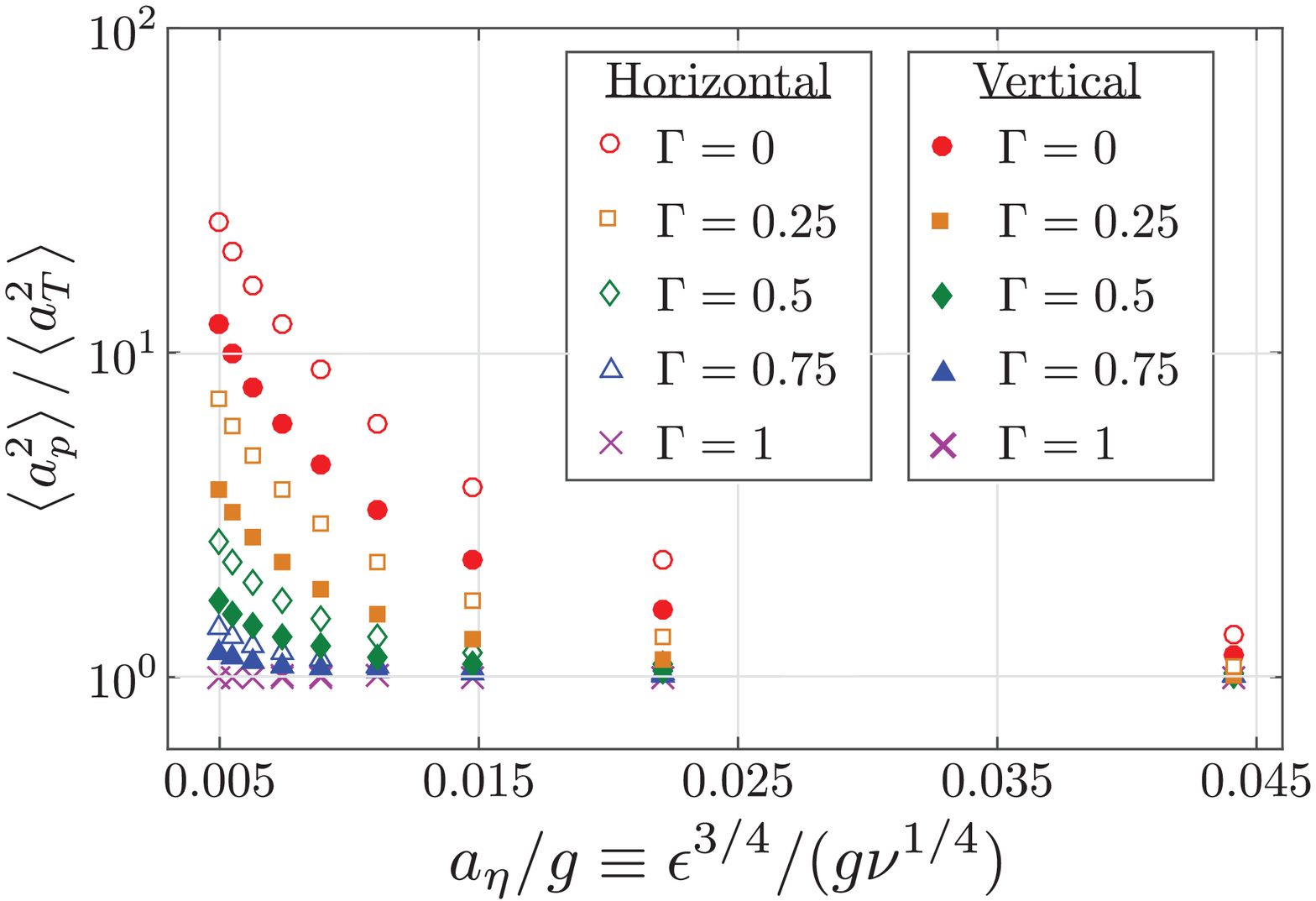}
\caption{Normalized acceleration variance for buoyant particles vs  $a_\eta/g$ obtained from Eulerian-Lagrangian DNS at Re$_{\lambda} \approx 80$. Hollow and solid symbols correspond to horizontal and vertical components, respectively.}
%Tiny bubbles~($D\approx 150 \pm 25 \mu$m) are generated from porous plates using pressurized air. The bubbles are illuminated using an arrangement consisting of a high-speed laser and volumetric optics. The camera is mounted on an arm which moves at constant velocity matching the flow speed. This ensures long particles trajectories. The particle size in the schematic has been greatly exaggerated for ease of comprehension. }
\label{fig:variance_vsFroude_numerics}
\end{figure}

We first consider the effect of changing the ratio of turbulent to gravitational acceleration, i.e $a_\eta/g$, which, according to~\cite{bec2014gravity}, equals the Froude number. Gravity enhances the acceleration in both vertical and horizontal directions (Fig.~\ref{fig:variance_vsFroude_numerics}). The vertical acceleration is consistently lower compared  to the horizontal, in agreement with our experiments. This is accompanied by a decrease in correlation time~(Fig.~\ref{fig:tcorr_vsFroude_numerics}), as also evidenced by our experiments.

\section{Non-dimensionalization}

When the advecting flow is turbulent and the particle spatial extension is of the order of the dissipative scale of turbulence, it is appropriate to non-dimensionalize eq.~(\ref{model_equation}) with respect to the Kolmogorov units: $\eta$ (length) and $\tau_{\eta}$ (time).
This leads to 
%\begin{equation}
%\ddot{\bfx} = \beta \frac{D\bfu}{D t}  + \frac{1}{\tau_p}(\bfu - \dot{\bfx}) + (\beta - 1) g \hat{\bfz}
%\end{equation}
%In dimensionless units
\begin{equation}
\ddot{\bfx} = \beta \frac{D\bfu}{D t}   + \frac{1}{\text{St}}(\bfu -  \dot{\bfx}) + \frac{1}{\text{Fr}} \hat{\bfe}_{z},
\label{eq_non-dim}
\end{equation}
where small case letters denote the new dimensionless variables and 
the following control parameters have been defined: 
\begin{equation}
\beta \equiv  \frac{3 \rho_f}{\rho_f + 2 \rho_p}; \quad \text{St} \equiv  \frac{d_p^2}{12 \beta \nu \tau_{\eta}} = \frac{\tau_p }{ \tau_{\eta}} ; \quad \text{Fr} \equiv  \frac{ a_{\eta}  }{ \left( \beta - 1 \right) g }
\label{StFrbeta}
\end{equation}
Note that the Stokes number St is defined taking added mass into account, and the Froude number Fr is a modified one that takes the particle density, through $\beta$, into account. Therefore these definitions are valid for a particle of arbitrary density.

In the high turbulence intensity limit ($\text{Fr} \to \infty$), the third right-hand-side term in the equation of motion can be neglected. 
Under this condition the vanishing St limit leads to $\dot{\bfx} \simeq \bfu$ for the velocity, and for the acceleration to $\ddot{\bfx} \simeq D_t \bfu$, where $D_t \bfu$ denotes the fluid-tracer acceleration.\\
At finite $\text{Fr}$, the small  St limit leads to $\dot{\bfx} \simeq \bfu+\frac{\text{St}}{\text{Fr}}\hat{\bfe}_{z}$.
For the particle acceleration this implies:
\begin{eqnarray}
%\ddot{\bfx} \simeq \frac{d}{dt}( \bfu + \frac{\text{St}}{\text{Fr}}\hat{\bfe}_{z}) 
%&=& \frac{\partial \bfu }{\partial t} + \dot{\bfx} \cdot \nabla ( \bfu + \frac{\text{St}}{\text{Fr}}\hat{\bfe}_{z})\\
%&=&  D_t \bfu+  \frac{\text{St}}{\text{Fr}}\hat{\bfe}_{z}  \cdot \nabla \bfu \\
\ddot{\bfx} \simeq  D_t \bfu + \frac{\text{St}}{\text{Fr}} \partial_z \bfu
\label{eq_accel}
\end{eqnarray}

\section{acceleration variance}
We consider the single-component acceleration variance. These are
%It is here convenient to distinguish between its horizontal ($x$ or $y$) and vertical cartesian components ($z$).
\begin{eqnarray}
\langle \ddot{x}^2  \rangle &\simeq& \langle (D_t u_x)^2  \rangle + \left( \frac{\text{St}}{\text{Fr}}\right)^2  \langle(\partial_z u_x)^2 \rangle,\\
%\langle \ddot{z}^2  \rangle &\simeq& \langle (D_t u_z)^2  \rangle + \left( \frac{\text{St}}{\text{Fr}}\right)^2  \langle(\partial_z u_z)^2 \rangle,
%\beta \langle \left( \frac{D}{D t} \bfu \right) ^2  \rangle + \frac{1}{St}(\bfu -  \dot{\bfx}) + \frac{1}{Fr} \hat{\bfz}
\end{eqnarray}
where $x$ and $z$ are the horizontal and vertical components, respectively. Note that the linear terms in $\text{St/Fr}$ vanish because there is no correlation between terms of the type $\bfu \cdot \nabla u_i$ and $\partial_z u_i$, i.e. there is no instantaneous correlation between the velocity field and its gradient.

Under isotropic turbulent conditions, the following relations are verified~\cite{hinze19720}:
\begin{eqnarray}
\langle(\partial_Z U_X)^2 \rangle \simeq \frac{2}{15} \frac{\epsilon}{\nu}, \\
\qquad \langle(\partial_Z U_Z)^2 \rangle \simeq \frac{1}{15} \frac{\epsilon}{\nu},\\
\qquad \langle (D_{\mathcal{T}} U_i)^2   \rangle = a_0 \epsilon^{3/2} \nu^{-1/2},
\end{eqnarray} 
where $i$ denotes one of the components x, y, or z, and $a_0$ is the so-called Heisenberg-Yaglom constant.
%And taking into account isotropy also implies that  
%$\langle(\partial_Z U_X)^2 \rangle \simeq \frac{2}{15} \frac{\epsilon}{\nu} $ and $\langle(\partial_Z U_Z)^2 \rangle \simeq \frac{1}{15} \frac{\epsilon}{\nu} $
%$\langle(\partial_z u_x)^2 \rangle \simeq \frac{2}{15} \frac{\epsilon}{\nu} $ and $\langle(\partial_z u_z)^2 \rangle \simeq \frac{1}{15} \frac{\epsilon}{\nu} $. 
From this, one obtains the relations linking the acceleration variance of particles to that of fluid tracers 
\begin{eqnarray}
\frac{ \langle \ddot{x}^2  \rangle}{ \langle (D_t u_x)^2  \rangle} &\simeq& 1 +  \frac{2}{15 a_0} \left( \frac{\text{St}}{\text{Fr}}\right)^2\\
\frac{ \langle \ddot{z}^2  \rangle}{ \langle (D_t u_x)^2  \rangle} &\simeq& 1 +  \frac{1}{15 a_0} \left( \frac{\text{St}}{\text{Fr}}\right)^2
\end{eqnarray}

These predictions are applicable to both heavy~($\text{St/Fr} <1$) and light~($\text{St/Fr}>1$) particles of arbitrary density. On the experimental side, we have confirmed the enhancement of acceleration variance using tiny air-bubbles dispersed in our water tunnel facility. For heavy-particles, our predictions remain to be experimentally verified.

%\begin{equation}
%\mathcal{F}( \ddot{x}) = \frac{\mathcal{F}(D_t u_x} + \mathcal{F}(\partial_z u_x)}{\partial_z u_x}
%\end{equation}

\section{Lagrangian time correlation}

\begin{figure} [!htbp]
	\centering
	\includegraphics[width=1\linewidth]{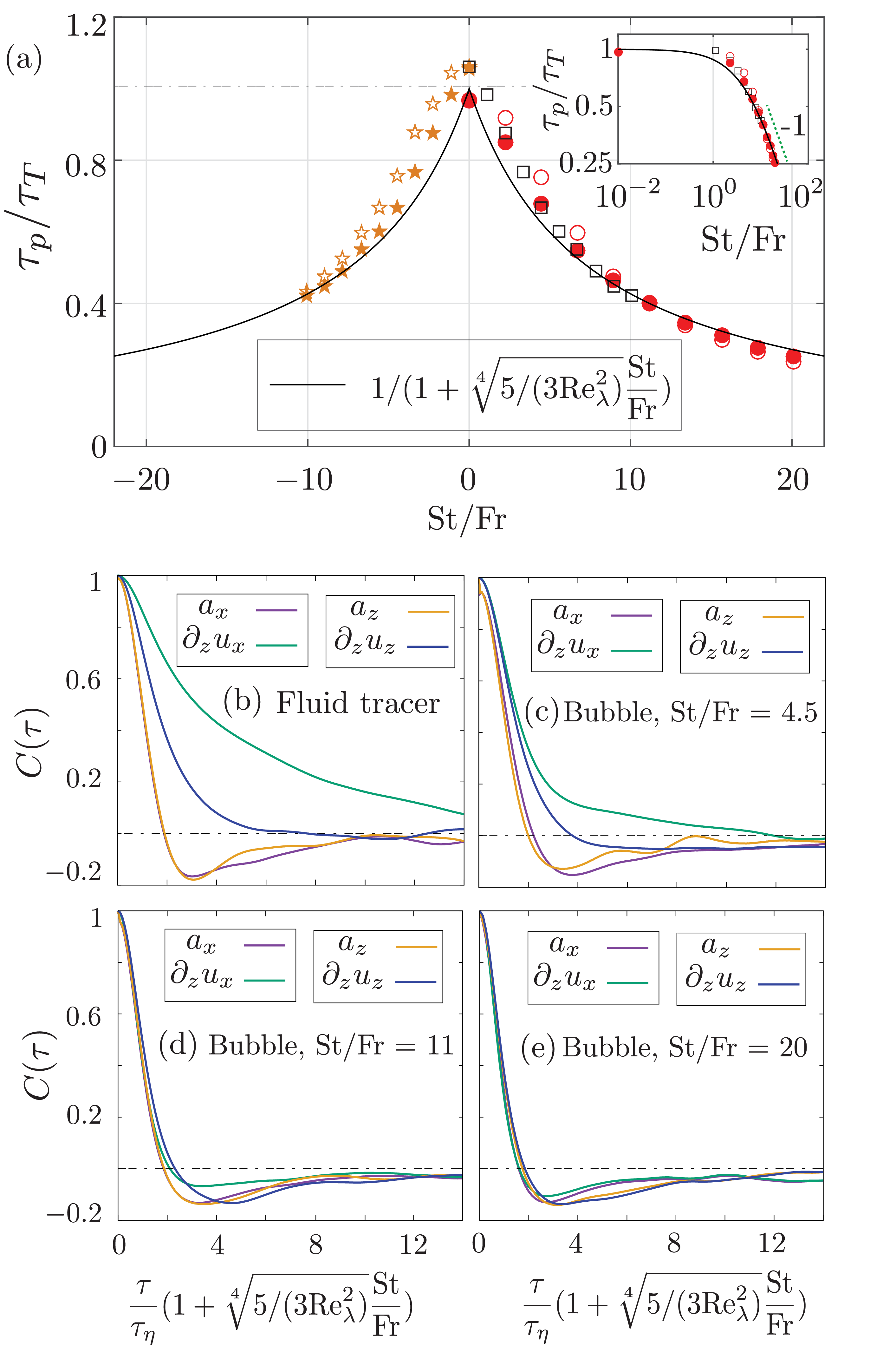}
	\caption{\textcolor{black}{(a)~Normalized correlation time for bubbles~($\beta = 3$) and very heavy~($\beta =0$) particles vs  St/Fr obtained from Eulerian-Lagrangian DNS at Re$_{\lambda} \approx 75$. The curve in black color shows the theoretical prediction, and the inset shows the plot on log-log scale. (b)-(e) Normalized autocorrelation function for fluid tracers and bubbles at different St/Fr values. Here, $a_x$ and $a_z$ are the horizontal and vertical accelerations, respectively. $\partial_z u_x$ and $\partial_z u_z$ are the spatial-velocitiy gradients, which contribute to the $a_x$ and $a_z$, respectively.  In (c)-(e) St/Fr increases from 4.5 to 20.}}
	%Tiny bubbles~($D\approx 150 \pm 25 \mu$m) are generated from porous plates using pressurized air. The bubbles are illuminated using an arrangement consisting of a high-speed laser and volumetric optics. The camera is mounted on an arm which moves at constant velocity matching the flow speed. This ensures long particles trajectories. The particle size in the schematic has been greatly exaggerated for ease of comprehension. }
	\label{fig:tcorr_vsFroude_numerics}
\end{figure}

\textcolor{black}{In Fig.~\ref{fig:tcorr_vsFroude_numerics}(a), we plot the simulation results for the evolution of Lagrangian time-correlation of acceleration with the ratio St/Fr. The left branch points to heavy particle, and the right one, to light particles. With increasing magnitude of St/Fr, we observe a decline in the correlation time for both heavy and light particles. Clearly, the drifting of the buoyant or heavy particle through the flow affects the correlation time. We model this by considering the case of a particle drifting through the flow at a speed $u_r$. In the absence of particle drift, it is well-known that the decorrelation time is $\tau_T \sim \tau_\eta$~\cite{yeung1989lagrangian}. Based on the characteristic velocity of a particle in the turbulent flow, $u_{rms}$, we estimate the length scale corresponding to this decorrelation time as $\Lambda \sim u_{rms} \tau_\eta$. Now, for a buoyant or heavy particle, the time of correlation is reduced due to an extra drift speed. Therefore, the new correlation time may be written as $\tau_p \approx \tau_T/(1+ u_r/u_{rms})$. For homogeneous isotropic turbulence the $u_{rms}$ may be expressed in terms of the Re$_{\lambda}$ and the $u_\eta$. This leaves us with the expression: $\tau_p/\tau_\eta \approx {1}/{(1+ \sqrt[4]{5/(3 \text{Re}_{\lambda}^2})\frac{\text{St}}{\text{Fr}})}$. The predictions, shown by the solid black curve, are in reasonable agreement with our numerical observations.}

\textcolor{black}{While the model provides reasonable predictions for the decorrelation time, we observe some small deviations from small to moderate St/Fr values in Fig.~\ref{fig:tcorr_vsFroude_numerics}(a). Below, we provide an explanation for these deviations. From eq.~(8), we note that the acceleration of a drifting particle has two contributions: (a) $D_t \bf u$ from the fluid tracer acceleration, and (b) $\frac{\text{St}}{\text{Fr}}\partial_z \bf u$ from the velocity-gradients in the flow. In Fig.~\ref{fig:tcorr_vsFroude_numerics}(b)-(e), we show the normalized time correlation of the particle accelerations and the gradient terms along the particle trajectories. For small St/Fr, the velocity gradient terms ($\partial_z u_z$ and $\partial_z u_x$) decorrelate slower as compared to the fluid acceleration term~(see Fig.~\ref{fig:tcorr_vsFroude_numerics}(b) \& (c)). Since St/Fr is small, the fluid acceleration term $D_t \bf u$ dominates over the velocity gradient term $\frac{\text{St}}{\text{Fr}}\partial_z \bf u$. This explains the slower decrease in the decorrelation time in the numerics as compared to our predictions for small St/Fr (see $\mid$St/Fr$\mid < 5$ in Fig.~\ref{fig:tcorr_vsFroude_numerics}(a)).  Some additional observations may be made for the moderate St/Fr range. From Fig.~\ref{fig:tcorr_vsFroude_numerics}(a), we note that the vertical component of particle acceleration (solid symbols) is shorter-correlated compared to the horizontal component~(hollow symbols). This occurs because the fluid velocity gradient components are not all correlated in the same way. Due to the incompressibility constraint ($\partial_i u_i = 0$), the longitudinal gradient $\partial_z u_z$ is shorter-correlated compared to the transverse gradient $\partial_z u_x$, as may be seen in Fig.~\ref{fig:tcorr_vsFroude_numerics}(b) \& (c). Since the vertical acceleration is influenced by the longitudinal velocity-gradient, it decorrelates in shorter time than the horizontal acceleration. We now consider the case of large St/Fr in Fig.~\ref{fig:tcorr_vsFroude_numerics}(a). In this case, the velocity gradient term $\frac{\text{St}}{\text{Fr}} \partial_z \bf u$ in eq.~(8) dominates over the fluid acceleration term $D_t \bf u$, and therefore, the decrease in correlation time in DNS is in good agreement with the predictions of our eddy-crossing model.}

\section{Acceleration Intermittency}
Intermittency, i.e. the observed strong deviations from Gaussianity, can be characterized in terms of the flatness of acceleration~$\mathcal{F}(a_p) \equiv \left <a_p^4 \right >/\left < a_p^2 \right >^2$. \textcolor{black}{Assuming statistical independence between $D_t u_i$ and $\partial_z u_i$\footnote{The absence of correlations between $D_t u_i$ and $\partial_z u_i$ is a first order~(to be refined) approximation.}, we obtain the tracer-normalized flatness of particle acceleration,} 

\small
\begin{align}
\frac{\mathcal{F}( \ddot{x})}{\mathcal{F}(D_t u_x)} = \frac{ 1 + \frac{12}{15\ a_0 \mathcal{F}(D_t u_x)}  \left( \frac{\text{St}}{\text{Fr}}\right)^2(1  +  \frac{\mathcal{F}(\partial_z u_x)}{45\ a_0}  \left( \frac{\text{St}}{\text{Fr}}\right)^2)}{ 1+ 2 \frac{2}{15\ a_0}  \left( \frac{\text{St}}{\text{Fr}}\right)^2  +  \left(\frac{2}{15\ a_0}\right)^2  \left( \frac{\text{St}}{\text{Fr}}\right)^4 }\\
\frac{\mathcal{F}( \ddot{z})}{\mathcal{F}(D_t u_x)} = \frac{ 1 + \frac{6}{15\ a_0 \mathcal{F}(D_t u_x)}  \left( \frac{\text{St}}{\text{Fr}}\right)^2(1  +  \frac{\mathcal{F}(\partial_z u_z)}{90\ a_0}  \left( \frac{\text{St}}{\text{Fr}}\right)^2)}{ 1+ 2 \frac{1}{15\ a_0}  \left( \frac{\text{St}}{\text{Fr}}\right)^2  +  \left(\frac{1}{15\ a_0}\right)^2  \left( \frac{\text{St}}{\text{Fr}}\right)^4 }
\label{flatness_norm}
\end{align}
\textcolor{black}{In the limit of small $\text{St/Fr}$,}
\begin{eqnarray}
\frac{ \mathcal{F}( \ddot{x})}{   \mathcal{F}(D_t u_x) } &\simeq&  1 - \frac{4}{15 \ a_0} \left(1 -  \frac{3}{ \mathcal{F}(D_t u_x)}  \right)  \left( \frac{\text{St}}{\text{Fr}}\right)^2\\
\frac{ \mathcal{F}( \ddot{z})}{   \mathcal{F}(D_t u_x) } &\simeq&  1 - \frac{2}{15 \ a_0}\left(1 - \frac{3}{ \mathcal{F}(D_t u_x)}  \right)  \left( \frac{\text{St}}{\text{Fr}}\right)^2
\label{flatness_theory_smallStbyFr}
\end{eqnarray}
It is verified that $\mathcal{F}(D_t u_x) > 3$. Therefore, both components are decreasing functions of  $\text{St/Fr}$, with the vertical being larger than the horizontal one for small St/Fr.\\
In the large $\text{St/Fr}$ limit,
\begin{eqnarray}
\frac{ \mathcal{F}( \ddot{x})}{   \mathcal{F}(D_t u_x) } &\simeq&  \frac{ \mathcal{F}(\partial_z u_x)}{   \mathcal{F}(D_t u_x) }\\
\frac{ \mathcal{F}( \ddot{z})}{   \mathcal{F}(D_t u_x) } &\simeq&  \frac{ \mathcal{F}(\partial_z u_z)}{   \mathcal{F}(D_t u_x) }
\label{flatness_theory_largeStbyFr}
\end{eqnarray}
%It is well known that $ \mathcal{F}(D_t u_x)  >  \mathcal{F}(\partial_z u_x) >\mathcal{F}(\partial_z u_z) $~\cite{ishiharaJFM2007}, meaning that particle rise/fall leads to a reduction in  flatness factor, which asymptotically approaches eq.~\ref{flatness_theory_largeStbyFr}. It is also verified that $\mathcal{F}(\partial_z u_x) >  \mathcal{F}(\partial_z u_z) $, which leads to the result $\mathcal{F}( \ddot{x}) > \mathcal{F}( \ddot{z})$. 

\textcolor{black}{It is verified that $ \mathcal{F}(\partial_z u_z) < \mathcal{F}(\partial_z u_x) < \mathcal{F}(D_t u_x)$~\cite{ishihara2007small}. This leads to the prediction $\mathcal{F}( a_v) < \mathcal{F}( a_h)$ i.e. vertical acceleration is less intermittent compared to the horizontal. }
	
\section{Interpretation}
	
\begin{figure} [!htbp]
	\centering
	\includegraphics[width=1\linewidth]{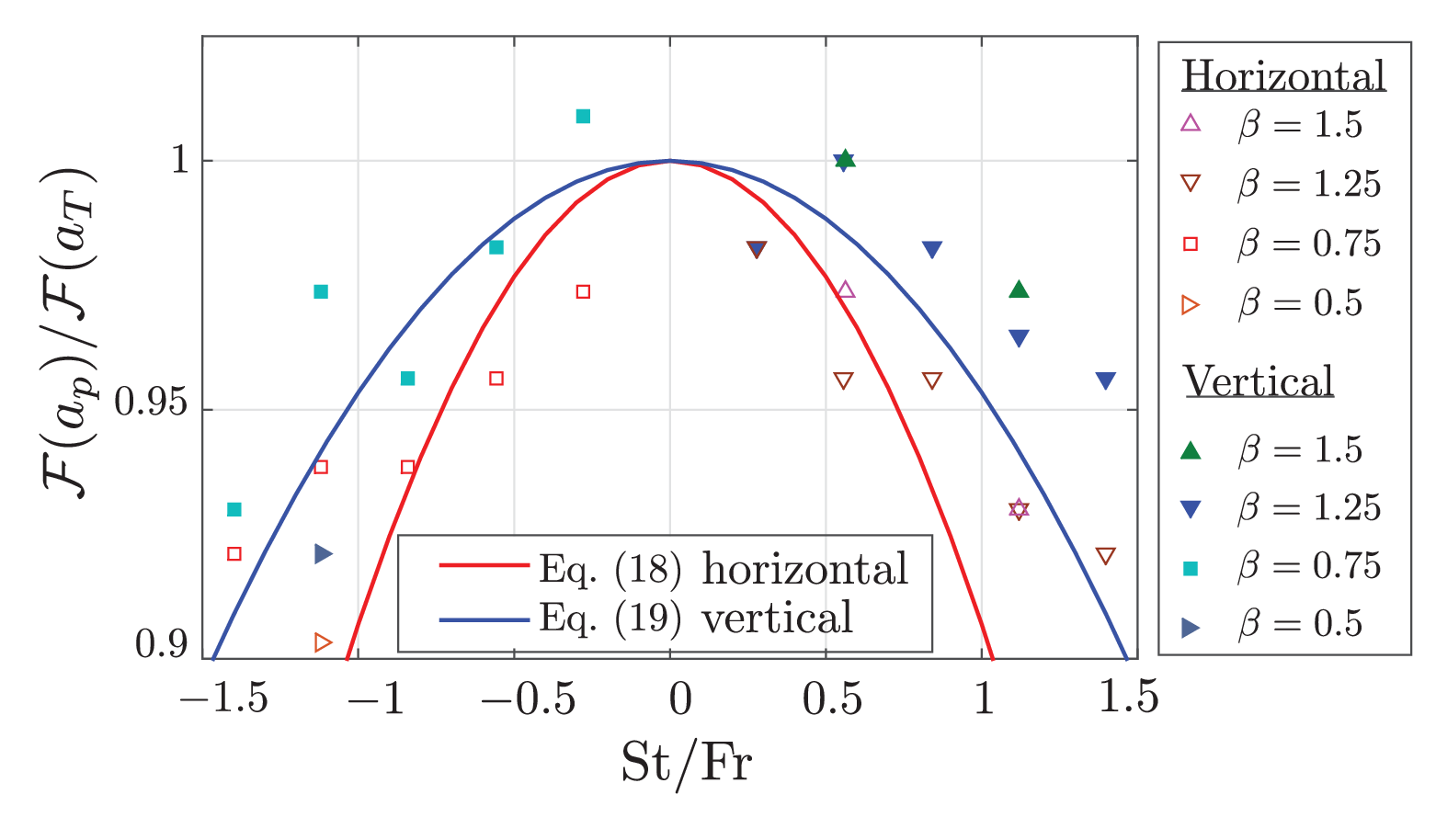}
	\caption{\textcolor{black}{Normalized Flatness factor in the small St/Fr limit. Hollow symbols show the flatness fo horizontal acceleration. Solid symbols show the flatness of vertical acceleration. The vertical acceleration flatness is mostly higher in numerics, in qualitative agreement with the predictions of eq. (18)-(19).}}
	\label{fig:Flatness_smallStokes}
\end{figure}

	\textcolor{black}{Eq.~(18)-(19) and eq.~(20)-(21) provide predictions for the normalized acceleration flatness~(intermittency) in the limits of small St/Fr and large St/Fr, respectively. Eq. (18)-(19) predict that the vertical component of flatness exceeds the horizontal one in the small St/Fr limit, while eq.~(20)-(21) predict that the horizontal component exceeds the vertical one when St/Fr is large. Therefore, an interesting cross-over is predicted between the flatness factors of the two components as one moves from small St/Fr to large St/Fr. In Fig.~\ref{fig:Flatness_smallStokes}, we present our numerical results in the small St/Fr range ($\text{St/Fr} < 1$). Despite the scatter in data, we \textcolor{black}{make} some interesting observation about our numerical results. With the exception of a single datapoint, the vertical components (solid symbols) are always higher compared to the horizontal components (hollow symbols), in agreement with eq.~(18)-(19). In the large St/Fr limit, as was clear from Figure 4(c) of the main paper, the horizontal component exceeded the vertical one, again in agreement with our predictions~(eq.~(20)-(21)). Therefore, the cross-over predicted by us is qualitatively seen in our simulations as well. A quantitative agreement is missing since the higher moments (Flatness) are in general very sensitive. Moreover, the lowest Stokes numbers possible in simulations is $\approx$ 0.05, which is still an approximation of the St $\to$ 0 limit. In addition, the predictions are subject \textcolor{black}{to a few assumptions}, such as the absence of correlation between velocity field and its gradient. While this is reasonable, it is not an exact result. At present, we do not have the resolution to verify the intricate details of (18)-(19). These aspects may be tested in future studies.}

\section{A generalized approach}
\textcolor{black}{We discuss some of these recent analytical and numerical approaches for heavy particles. 
% As reported by \citet{aliseda2002effect}, most of the progress in this topic has come from numerical simulations of homogeneous isotropic flow~\cite{bec2014gravity,parishani2015effects,ireland2015effect}, with little support from the experimental side.
 Interesting effects have been demonstrated on the gravity-induced modification of heavy particle statistics~\cite{bec2014gravity,gustavsson2014clustering, parishani2015effects,aliseda2002effect,ireland2015effect}, which were quantified as a function of particle inertia (St) and the Froude number given by the ratio of turbulent to gravitational acceleration ($a_\eta/g$).}
\textcolor{black}{We examine the Stokes and Froude number definitions used in these studies. The Stokes number was defined as $\text{St} \equiv \frac{\rho_p d_p^2}{18 \rho_f \nu \tau_\eta}$~\cite{bec2014gravity,parishani2015effects,ireland2015effect}. This definition is appropriate at large $\rho_p/\rho_f$, when added mass effects are negligible. We also note that the definition of Froude number as $a_\eta/g$~\cite{bec2014gravity,ireland2015effect} does not take the particle density into account. Here, $a_\eta$ is the turbulent acceleration and $g$ is the gravitational acceleration. This Froude number definition is applicable when the particles under consideration are of fixed $\rho_p/\rho_f$. Therefore, these results apply to the case of very heavy particles and when the density ratio is kept constant, i.e $\rho_p/\rho_f$~=~constant~$\gg 1$.} 

\textcolor{black}{In this paper (main article), we provide a generalized description that is applicable to particles of arbitrary density. We have numerically shown the validity of our theoretical preditions for particles of arbitrary density. The theory we develop for isotropic turbulence can explain also our experimental findings on bubbles in a turbulent water flow.  The generic St and Fr definitions we use converge to the definitions in~\cite{bec2014gravity,parishani2015effects,ireland2015effect} at large $\rho_p/\rho_f$. Therefore, \textcolor{black}{such a modified approach may be useful} for future studies that explore the effects of gravity for arbitrary-density particles in turbulence.}
\bibliography{mybib_new_f}